# The Evolving Activity of the Dynamically Young Comet C/2009 P1 (Garradd)

D. Bodewits[1*], T. L. Farnham[1], M. F. A'Hearn[1], L. M. Feaga[1], A. McKay[2], D.G. Schleicher[3], and J. M. Sunshine[1]

[1] Department of Astronomy, U. Maryland, College Park, MD 20742-2421 USA; [2] Astronomy Department, New Mexico State University; [3]Lowell Observatory, 1400 West Mars Hill Road, Flagstaff, AZ 86001; * corresponding author: dennis@astro.umd.edu,

*Keywords:* comets: general – comets: individual (C/2009 P1 Garradd) – Oort Cloud – techniques: image processing – techniques: photometric – ultraviolet: planetary systems

## Abstract

We used the UltraViolet-Optical Telescope on board Swift to observe the dynamically young comet C/2009 P1 (Garradd) from a heliocentric distance of 3.5 AU pre-perihelion until 4.0 AU outbound. At 3.5 AU pre-perihelion, comet Garradd had one of the highest dust-to-gas ratios ever observed, matched only by comet Hale-Bopp. The evolving morphology of the dust in its coma suggests an outburst that ended around 2.2 AU pre-perihelion. Comparing slit-based measurements and observations acquired with larger fields of view indicated that between 3 AU and 2 AU pre-perihelion a significant extended source started producing water in the coma. We demonstrate that this source, which could be due to icy grains, disappeared quickly around perihelion. Water production by the nucleus may be attributed to a constantly active source of at least 75 km$^2$, estimated to be >20% of the surface. Based on our measurements, the comet lost $4 \times 10^{11}$ kg of ice and dust during this apparition, corresponding to at most a few meters of its surface. Even though this was likely not Garradd's first passage through the inner solar system, the activity of Garradd was complex and changed significantly during the time it was observed.

## 1.  INTRODUCTION

Comets are generally linked to the formation history of our solar system and several studies have tried to establish chemical taxonomies, as these might reflect the formation conditions in the protosolar disk (A'Hearn et al. 1995; Mumma & Charnley 2011; A'Hearn et al. 2012). It is unclear however which properties of comets are primordial and which are the product of subsequent evolution, which hampers our understanding of the connection between comets and the protosolar disk. Dynamically new comets, approaching the Sun for the first time, are known to behave differently from other classes of comets, whether due to inherent compositional differences among the various comet classes or to evolutionary effects of solar heating (Oort & Schmidt 1951; Whipple 1978; A'Hearn et al. 1995). New comets have resided in the Oort cloud, where ices that accumulated during their formation should have suffered cosmic ray processing of their surfaces producing shells of highly volatile radicals (Johnson et al. 1987; Stern et al. 2003). Activation and depletion of the different volatiles during the early approach to the Sun govern the activity as the comet evolves toward more typical behavior later in the apparition (Meech & Svoren 2005; Meech et al. 2009). Oort & Schmidt (1951) noted that new comets tend to be more active on their way towards the Sun. A'Hearn et al (1995) suggested this inbound hyper-activity evolves into more regular activity later in the apparition. It is unknown whether behavior like this is driven by comet evolution (the removal of outer layers) or that it reflects primordial heterogeneity reflecting comets origins within the solar nebula. Long-term monitoring of Oort Cloud comets is needed to help determine how the evolutionary processes differ from other classes of comet.

Comet C/2009 P1 (Garradd) was a bright, active comet, well observable over a wide range of heliocentric distances, and it is the first comet for which all three main volatiles ($H_2O$, $CO_2$, and CO) have been observed, or at least inferred, along a significant part of its passage through the inner solar system (Combi et al. 2013; Feaga et al. 2014; McKay et al. 2012; DeCock et al. 2013). Dynamical solutions indicate that comet (Garradd) has an original reciprocal semi-major axis of $(1/a)_0 = 0.000390 – 0.000403$ AU$^{-1}$ which suggests that this was not the first time it approached the Sun (S. Nakano, 2011[1]; Minor Planet Center[2]). Dynamically young comets like Garradd may have lost their crust of highly volatile radicals, but they may provide a link to the primordial, least-processed material that is now at or close to the surface. As such, dynamically young comets are likely in a transitional stage that can prove an invaluable key to how comets work, and on how Oort Cloud comets evolve.

In this paper we present space-borne observations acquired with the Ultraviolet-Optical Telescope on board the Swift spacecraft (Sec. 2). We describe our data analysis in Sec. 3 and present our results in Sec. 4. In Sec. 5 we then use our measurements to discuss how the activity of comet Garradd evolved along its passage through the inner solar system.

---

[1] NK 2109; http://www.oaa.gr.jp/~oaacs/nk/nk2109.htm
[2] http://www.minorplanetcenter.net/

## 2. OBSERVATIONS

### 2.1 Swift/UVOT Observations

Swift is a multi-wavelength observatory equipped for rapid follow-up of gamma-ray bursts (Gehrels et al. 2004). Its Ultraviolet-Optical Telescope (UVOT) has a 30 cm aperture that provides a 17 × 17 arcminute field of view, with a plate scale of 1 arcsecond/pixel and a point spread function of 2.5 arcseconds FWHM (Mason et al. 2004; Roming et al. 2005). UVOT is equipped with a photon counting detector. This results in very low background levels but has the disadvantage that it is limited at high incident fluxes due to coincidence loss, i.e. the arrival of more than one photon in a given pixel during a single readout of the detector ('coincidence loss'; Kuin & Rosen 2008; Breeveld et al. 2010). Seven broadband filters allow color discrimination, and two grisms provide low-resolution spectroscopy at UV and optical wavelengths (1700–6500 Å). These grisms provide a resolving power $R = \lambda/\Delta\lambda \sim 100$ for point sources. In optical and UV wavelengths, the cometary spectrum consists of sunlight reflected by ice or dust grains in the coma, and overlying emission features of molecules and ions. Swift/UVOT is not equipped with narrowband filters commonly used for cometary studies. We therefore used the UVW1 (central wavelength $\lambda_c$ 2600 Å, FWHM 700 Å) and V-band ($\lambda_c$ 547 Å, FWHM 750 Å) filters to measure the number of OH molecules in the field of view, and to estimate C/2009 P1's water and dust production rate during our observations. Inevitably, the flux measured by these broadband filters includes both reflected continuum and emission lines. We will discuss these contaminations in Sec. 3.

Swift/UVOT acquired 107 observations of comet C/2009 P1 (Garradd) between April 2011 and October 2012, covering a range of heliocentric distances of 3.35 AU to 1.55 AU inbound, and obtained its last observations of the comet post-perihelion at a distance of 4.0 AU from the Sun. UVOT cannot observe targets at solar elongations of less than 50 degrees and did not monitor the comet between Nov. 14 – Dec. 25, 2011 and June 1 – Oct. 15, 2012. The data were processed through the standard Swift/UVOT pipeline which delivers cleaned, calibrated images to the HEASARC archive (Roming et al. 2005; Breeveld et al. 2010). The observing log is summarized in Table 1 and a sample of the images is shown in Fig. 1.

Comet Garradd has a highly inclined orbit ($i = 106$ deg) and its observing geometry changed significantly during the time we observed it. During the first three observations, the comet was observed with a very low phase angle, with its tail behind it as seen from Earth. Garradd crossed the ecliptic plane around $r_h = 3.4$ AU pre-perihelion. During the observations at $r_h = 2.0$ AU inbound, the Earth was far above the comet's orbital plane, revealing a broad dust tail. The comet reached its perihelion ($r_h = 1.551$ AU) on Dec. 23.55, 2011 UT. After this, at $r_h = 1.7$ AU post perihelion, the comet and Earth reached a similar observing geometry where Earth was far above the orbital plane. The comet crossed the ecliptic plane again around $r_h = 3.0$ AU while it was behind the Sun as seen from Earth. Our last Swift observations were acquired on Oct. 21, 2012 UT, at a heliocentric distance of 3.99 AU and 4.48 AU from Earth. The differences in observing geometry mainly affect how much of the tail is visible.

## 3. ANALYSIS

### 3.1 Photometry and Morphology

Comae are very extended clouds of gas and grains, and UVOT's field of view (17 x 17 arcminutes) allows us to cover a larger area than most spectroscopic instruments. The drawback of this is that every image inevitably contains a large number of background objects that can significantly add to the measured flux. To remove these we produced azimuthal median surface brightness profiles of the coma. This was achieved by converting each UVOT image into polar coordinates and finding the resistant mean value of the surface brightness B(r) at a given radial distance r. To derive water production rates, we used these surface brightness profiles to measure the fluxes in the UVW1 (which covers the OH emission around 3000 Å) and V (for continuum) bands. Apertures of 50 arcseconds in radius were used to cover a large fraction of the coma and achieve good signal-to-noise. This corresponded to a projected radius between $0.55 - 1.6 \times 10^5$ km at the comet. For the first two sets of observations pre-perihelion and the last observation post-perihelion we stacked all individual exposures and used smaller apertures of 25 arcsec in radius to increase the SNR.

To derive Afρ, a proxy for the dust content of the coma (A'Hearn et al. 1984) we used V-band images. To circumvent aperture effects we fixed the aperture radius to $5 \times 10^4$ km. This resulted in apertures with radii between 15 – 54 arcseconds, which are large enough to warrant the use of our azimuthal profiles to measure the flux. For Afρ measurements, it is desirable to use smaller apertures in order to avoid processes that affect the dynamics or population of the dust grains. The smallest reasonable constant aperture would be $2 \times 10^4$ km, corresponding to 5 pixels in radius, comparable to UVOT's point spread function. However, within such small apertures centered on the optocenter coincidence loss becomes a significant problem that is not easily corrected for in extended sources, underestimating count rates in the central 5-pixel aperture by at least a factor 2 (Poole et al. 2008). We minimized the effect of coincidence loss on our measurements by using slightly larger apertures of $5 \times 10^4$ km in radius, corresponding to 15 – 54 arcseconds. This assured that the center 5 pixels did not contribute more than a few percent to the total measured flux.

Background count rates for both filters were determined from the outer parts of the CCD and are of the order 0.002 counts $s^{-1}$ pixel$^{-1}$ (UVW1) and 0.02 counts $s^{-1}$ pixel$^{-1}$ (V-band).

To study the dust morphology V-band images were divided by the azimuthal median surface brightness profiles described above (Sec. 5.3).

### 3.2 Continuum Removal

At the wavelengths covered by UVOT, comets are seen in sunlight reflected by cometary dust, with several bright molecular emission bands superimposed. The UVW1 filter is well placed to observe very strong 1-0, 0-0, and 1-1 bands of the OH [A $^2\Sigma$ – X $^2\Pi$] electronic transition at 2811, 3064, and 3122 Å, respectively. We convolved the UVW1 and V-band filter transmissions with an un-reddened solar spectrum to determine how much the continuum contributes to the UVW1 flux. We further assume that the flux in the V-band filter is dominated by continuum emission (while in truth it is contaminated by the fluorescent emission of various molecules, predominantly $C_2$ and $NH_2$; see Sec. 3.4). The OH

flux is then given by:

$$F_{OH} = \alpha \cdot (F_{uvw1} - \beta F_v)$$

where $F_{uvw1}$ and $F_v$ are the fluxes measured in the UVW1 and V-band filters, $\alpha$ is the transmission of the UVW1 filter at the wavelength of the OH transitions ($\alpha \sim 0.5$), and β is the ratio of continuum fluxes as measured with the two filters (β = 0.134 for the solar spectrum). Using this relation, we find that the continuum contributed between 25– 87% of the flux measured with the UVW1 filter, depending on the comet's dust-to-gas ratio and the variable fluorescence efficiency of OH. To investigate the effect of reddening on the continuum subtraction from the UVW1 images we multiplied solar spectra with different reddening slopes (normalized at 4030 Å, the midpoint between the UVW1 and V-band filters). At typical levels of reddening, 15% per 100 nm, the ratio between the solar flux measured in the UVW1 and V-band becomes β = 0.1.

For the observation on Oct. 14, 2012 UT ($r_h$ = 3.92 AU, Δ = 4.50 AU) no simultaneous V-band observation was acquired. For this observation we therefore give an upper limit for the water production rate directly based on the flux measured in the UVW1 filter. For the observations acquired on Oct. 21, 2012 UT ($r_h$ = 3.99 AU, Δ = 4.48 AU) we stacked the two UVW1 and V-band images (acquired in UVW1-V band pairs, 6.4h apart) to increase the signal-to-noise, yet found no evidence of residual flux. We calculated 3-sigma upper limits of the water production rate by propagating the stochastic uncertainty.

### 3.3 Gas Production Rates and Dust Content

The fluxes measured by Swift are listed in Table 1. To derive water production rates from the photometry, we first derived the total number of OH molecules within the field of view. We calculated fluorescence efficiencies for the three OH [A $^2\Sigma$ – X $^2\Pi$] bands in the band pass, accounting for the variation with the comet's heliocentric velocity (Schleicher & A'Hearn 1988), and assuming a $r_h^{-2}$ decrease with heliocentric distance. We compared the measured OH content of the coma with an OH distribution calculated using the vectorial model (Festou 1981; Combi et al. 2004). In brief, the vectorial model assumes an isotropic expansion of parent species assuming a fixed velocity. When these molecules dissociate, their fragments get a velocity kick and are accelerated. We assumed a water lifetime of 8.6 × 10$^4$ s, an OH lifetime of 1.29 × 10$^5$ s (both at 1 AU and scaled for larger heliocentric distances by $r_h^{-2}$), a bulk outflow gas velocity of 0.85 × $r_h^{-½}$ km s$^{-1}$ , and a constant OH velocity of 1.05 km s$^{-1}$ (Combi et al. 2004).

We used V-band photometry to derive values of Afρ, using a value of $m_v$ = -26.74 calculated for the brightness of the Sun in the Swift V-band passband. To account for different scattering efficiencies of dust grains observed under different phase angles we applied the phase function derived by Schleicher et al. (1998), normalized to a phase angle of 0 degrees (Fig. 2).

### 3.4 Uncertainties

Comet Garradd was relatively bright and made an excellent target for Swift. As a result of the high count rates, stochastic errors are negligible (<1%) comparable to other uncertainties for most observations, except for the individual UVW1 images acquired around $r_h$ = 4 AU outbound, where they were of the order of 5%. Swift/UVOT is very well calibrated and the flux measurements are accurate to within 5% (Poole et al. 2008). Swift/UVOT uses a microchannel plate intensified CCD (Poole et al. 2008). UVOT is thus insensitive to cosmic rays but susceptible to coincidence loss at high photon fluxes (>10 counts/s). This can be reasonably corrected for bright point sources, but is not readily done for extended sources. In the case of most comets, significant coincidence loss is limited to the innermost pixels due to the dust profile. As discussed above, we avoided this problem by using larger, fixed-km apertures; when the comet was the brightest (and closest to Earth), the area suffering most from coincidence loss contributed less to the total flux measured within the aperture. Based on the published coincidence loss corrections, we estimate the contribution to the systematic uncertainty of the flux measurement to be less than 5% in the V-band and negligible for the UVW1 band measurements (where count rates are much lower).

The use of broadband filters to measure OH and continuum fluxes inevitably implies that our results are relatively crude measurements of the gas and dust content of the coma. The bandwidth of the UVW1 filter encompasses faint features of $C_2$, CS, and $CO_2^+$, but these are typically more than an order of magnitude fainter than the OH lines. The UVW1 filter has a significant red tail but as the transmission is only 7% at 3500 Å we consider contamination by CN and NH emission to be minor. The exact extent of these contaminations is hard to assess. We discussed the effect of reddening on the derived OH flux in Sec. 3.2. We assumed neutral dust colors here. This implies we are subtracting too much flux from the UVW1 images when the dust is redder in color. The relative contribution of the continuum to the flux measured in the UVW1 filter decreased from 87% to 25% when the comet approached the Sun, and the effect of reddening would thus decrease at smaller heliocentric distances.

Several bright emission features fall within the band pass of the V-band filter, most notably those of $C_2$ and $NH_2$. The V-band contains the bandhead of the $\Delta v$ = 0 Swan-band sequence of the $C_2$ molecule. The relative contribution of contaminants to the flux measurements with the V-band depends on the size of field of view, since gas and dust distributions differ, and also on the gas and dust content of the coma. A crude estimate of the contribution of the $C_2$ filter to V-band flux can be derived by assuming an average $C_2$/OH abundance of 1/300 (A'Hearn et al. 1995) and by scaling the fluorescence efficiencies of the $C_2$ $\Delta v$ = 0 band and the OH [A $^2\Sigma$ – X $^2\Pi$] bands (~250, Bodewits et al. 2011). The transmission of the V filter increases sharply between 5000 – 5200 Å, resulting in a net transmission of 50%. Combining those numbers, we estimate the flux from $C_2$ to contribute less than 20% to the flux measured in the V-band. At larger heliocentric distances, where McKay et al. (2013) suggest that the rapid change in the observed strength of $C_2$ emission is attributed to the onset of a parent other than $C_2H_2$, thus indicating that the $C_2$-to-dust ratio may in fact be lower, the contamination is likely considerably less.

Additional systematic uncertainties are introduced by the assumption needed to derive water production rates from the measured column densities. We therefore estimate the systematic uncertainty in the water production rates to be 25%.

## 4. RESULTS

The water production rates derived from the Swift/UVOT measurements are presented in Figure 2. Approaching the Sun, Garradd's water production rate increased steeply following an $r_h^{-6}$ relation. The water production rates peaked early, 200 days before perihelion, at a rate of $2 \times 10^{29}$ molecules/s. It then remained constant for about 100 days, after which it decreased with approximately $r_h^{-4}$. The water production is clearly asymmetric around perihelion.

We also compare our measurements with those acquired by others (Fig. 3). This data set encompasses a large variety of methods and wavelength regimes. It contains direct spectroscopic measurements of water emission in the near- and far-infrared (Paganini et al. 2012; Villanueva et al. 2012; Feaga et al. 2014; Disanti et al. 2014; Bockelée-Morvan et al. 2012). It also contains photometric observations of fluorescent OH emission (our Swift UVOT data, as well as Deep Impact/MRI and Lowell observations by Farnham et al., in prep.), as well as Far-UV observations of the Ly-$\alpha$ halo acquired by SOHO/SWAN (Combi et al. 2013). Most data appear in good agreement, but high-resolution spectroscopic measurements of $H_2O$ emission in the Near-IR acquired with apertures of projected width of <1000 km (Paganini et al. 2012; Villanueva et al. 2012; DiSanti et al 2014) present water production rates that were two or three times lower than those acquired with larger fields of view (with radii of order $10^5$ km). A comparative study by Combi et al. (2013) suggests a trend between water production rate and aperture size, which they attributed to an extended $H_2O$ source in the coma. We will discuss this further in Section 5.

The dust content of the coma as measured by A(0)f$\rho$ shows a complex relation with the comet's distance to the Sun (Fig. 2). For the most part, it is nearly constant from our first measurement at $t_p$ - 244 d, through $t_p$ - 70 d, though there is an interesting local minimum around $t_p$ -154 d. It shows a shallow decrease after the time of the peak water production rate. At large heliocentric distances where the water production is low, Garradd had an extreme $^{10}$log A(0)f$\rho$ /Q(OH) ratio of -24.6 (cm×s/molecule) before perihelion, and perhaps even higher post-perihelion. This is higher than all comets in the 85-comet survey by A'Hearn et al. (1995), comparable only with comet Hale-Bopp ($r_h$ = 1.02 AU) and C/1991B (Shoemaker-Levy) ($r_h$ = 3.04 AU; A'Hearn et al. 1995; Farnham et al. 1997; Fernandez 2000)(we apply a correction for phase effects that was not applied to the measurements presented by A'Hearn *et al.*, but even without this correction the dust-to-gas ratio is higher than the value measured for Hale-Bopp). However, when we compare the ratio around perihelion, log A($\theta$)f$\rho$/Q(OH) is -25.3 (cm×s/molecule), a factor of 16 lower than our first measurement at $t_p$ = -244 d. Garradd's perihelion distance of q = 1.55 AU places it right in the middle of the comet population in A'Hearn et al. (1995). Post-perihelion, A($\theta$)f$\rho$/Q(OH) increased again by as much as a factor of 4 around $t_p$ - 132 d, and perhaps by as much as another factor of 4 around $t_p$ – 302 d.

## 5. Discussion.

In this section we will compare our measurements with those presented in other studies of Garradd's gas and dust production rates. We will then use this to try to reconstruct the evolution of the comet's activity. An overview of the gas and dust production rates is shown in Fig. 3.

**5.1 Water Production from the Nucleus and in the Coma**

The water production rate of Garradd is characterized by four features. First, the water production rate had a much steeper relation with heliocentric distance pre-perihelion than it had when the comet receded from the Sun. Secondly, the production rates of water and dust shown in Fig. 3 all peak around $t_p$ - 100 to $t_p$ - 70 d. Thirdly, we noted a very large variation of the dust-to-gas ratio with heliocentric distance (Fig. 2). Fourth, per-perihelion, slit-based instruments observed significantly lower water production rates than large-aperture measurements. DiSanti et al. (2014) noted that the sunward excess of water emission observed in all IR observations pre-perihelion was not present in their observation just after perihelion (Jan. 18, 2012 UT), which attributed this to the disappearance of a halo of icy grains near the nucleus.

We modeled the comet's water production using the sublimation model by Cowan & A'Hearn (1979), assuming that every surface element has constant solar elevation (as would be the case if the rotational pole pointed at the sun) or if the nucleus was very slowly rotating) and is therefore in local, instantaneous equilibrium with sunlight. This maximizes the sublimation averaged over the surface. We further assumed a Bond albedo of 0.05 and 100% infrared emissivity[3]. Combining this model with the water production rates derived from the Swift measurements, we calculated the required minimum active area on the comet. The results are shown in Fig. 4. It is striking that comparing the earliest measurements of the water production rate ($t_p$ - 250 d), the near-nucleus water production rates measured at $t_p$ – 100 d (Paganini et al. 2012, Villanueva et al. 2012), and those obtained well beyond perihelion (> $t_p$ + 100 d) suggests a near constant active area of approximately 75 km². We interpret this as the nuclear component to the water production rate. This constant area of 75 km² would result in a water production rate of $2.5 \times 10^{27}$ molecules/s at $t_p$ + 300 d, in agreement with the upper limit of $2 \times 10^{27}$ mol./s derived from the Swift/UVOT measurements.

The constant active area corresponds to a minimal radius of 2.5 km if the entire surface were active. Boissier et al. (2013) found an upper limit of 5.6 km based on millimeter observations of the continuum emission from the nucleus, which suggest a continuous active surface fraction of >20%. This would be a significant part of Garradd's surface and much higher than the active areas deduced for most comets for which reliable size estimates exist. The relatively large active fraction (comparable in size to the comet's cross section) validates the assumption of our sublimation model, which yields the smallest active area of those discussed in the Cowan & A'Hearn (1979) paper. It is also consistent with the suggestion that dynamically young comets have more active nuclei than short period comets (Meech et al. 2004).

There is a clear increase in the computed water active area that peaks around $t_p$ – 130 d and ends around perihelion. This bump coincides with the difference observed between slit-based and large aperture observations. Based on the water sublimation model the active area responsible for the observed increase in the water production rate would have been at least 300 km² if it where produced by the nucleus, e.g. it would require the entire nucleus to be active. Given that comets have low thermal conductivity and Garradd rotates relatively slowly at 10.4 ± 0.05 h (Bodewits, Farnham, & A'Hearn, 2012; Farnham et al. 2013), we consider it more likely that there is an additional source of water in the coma, such as icy grains, suggested by several authors (Paganini et al. 2011; Combi et al. 2012; Bockelée-Morvan et al. 2012). This additional source of water peaked 100 days before perihelion when it contributed up to 75% of the total water production rate. Its contribution decreased rapidly afterwards. Based on extensive modeling of the spectral line map of $H_2O$ observed with Herschel, Bockelée-Morvan et al. (2014) suggest the presence of an extended $H_2O$ source at $t_p$ + 60d responsible for 30% of the total water production rate then, whereas our model suggest a contribution at the 10% level. Given the disappearance of the sunward excess reported by DiSanti et al. (2014), and given the assumptions made and the consequent uncertainties, we deem all results consistent.

---

[3] Available on http://pdssbn.astro.umd.edu/tools/software.shtml#analysis (version of 12/1/2013).

## 5.2 Dust Production and Morphology

A(0)fρ is only a crude measure of the dust content in the coma, intended originally to be only a way of comparing observations at different geometries assuming other parameters were constant. Estimating the mass of the dust produced requires many assumptions (see Weiler et al. 2003 and Fink & Rubin (2012) for broader discussions). A'Hearn et al. (1995) quoted (from Arpigny, unpublished) an empirical correlation with dust production rates measured in the infrared to show that 1 cm in A(0)fρ corresponds to ~1 kg/s of dust production. However, when observing a comet over a large range of heliocentric distances, one should take the decreasing outflow velocity of the bulk gas into account, and even its effect on the size of particles that can be dragged from the surface. The bulk velocity decreases with $r_h^{-\frac{1}{2}}$ (Delsemme, 1982). We therefore calculate the dust mass loss rate using the relation $Q_d$ = A(0)fρ/$r_h^{\frac{1}{2}}$ (Fig. 2 and Fig. 3). The resulting dust loss rate is distinctly shallower than the gas loss rate, resulting in a large variation of the observed gas-to-dust ratio relative to A(0)fρ. The gas-to-dust ratio thus appears to increase when the water production goes from coma-dominated to nucleus dominated, which is counter-intuitive at first sight; coma produced water would not drive dust grains into the coma. This overall trend is difficult to interpret without further knowledge of the dust properties (albedo, size distribution) and how these changed in time. It is interesting to note that a similar relation was observed in comet Hale-Bopp, albeit over a much longer time scale (from 5 AU pre-perihelion to 13 AU post-perihelion; Weiler et al. 2003).

There are however interesting aspects to the morphology and dust-to-gas trend that are worthy of some further consideration. Inspection of the dust tail suggests that Garradd may have experienced a short burst of activity at a large heliocentric distance as it approached the Sun. Figure 5 shows a series of images of the dust tail, enhanced by dividing out an azimuthally averaged radial profile to show local variations in the spatial density of the dust (green and then red designate higher relative densities). Synchrones are superimposed on the image to provide an estimate of when the dust in the various tail regions was emitted. The December 2011 and the March 2012 images reveal an enhancement in the density in the vicinity of the $t_p$ - 240 day synchrone, which is highlighted in red for reference. This enhancement is a localized high-density cloud of dust that indicates there was a temporary increase in dust production ~8 months before perihelion. Along later synchrones, the density decreases again, until the dust production begins to ramp up closer to perihelion. This same characteristic is present on earlier dates back to -154 days, though it is not as obvious due to the lack of separation of the synchrones.  The earliest image, from June 2011, was obtained around the time of the outburst, and therefore the dust has had little time to spread out along the synchrones (and is thus removed in the azimuthal averaged enhancement process.  By the time of the April image, the cloud has moved out the field of view. Unfortunately, the morphology and overlapping of synchrones for very early times 'pile up' and start to overlap, making it impossible to determine exactly when the activity started, but the change in morphology between $t_p$ - 202 and -154 days suggests that the burst ended ~6 months before perihelion.

The outburst scenario also provides an explanation for the anomalous behavior of Afρ in comet Garradd. As discussed earlier, measurements prior to $t_p$ - 180 days are elevated in comparison to the later measurements (Fig 4) showing the effects of the increase in dust production. The local minimum in Afρ at $t_p$ - 154 days is consistent with lower levels of dust after the cessation of the outburst. We also note that the increase in the dust-to-gas ratio after perihelion is likely due to CO emission.  Since this measurement only includes water in the gas term, any dust emitted as a result of sublimation of other gases will skew the dust-to-gas ratio. Feaga et al (2014) showed that CO production ramped up monotonically through perihelion, approaching the water production at $t_p$~100 days (Fig. 3).  This high production rate is undoubtedly entraining dust that elevates the dust-to-gas ratio measurement.

### 5.3 Mass Loss Erosion

Garradd was well observable throughout its passage through the inner solar system and we can derive a reasonable estimate of the amount of material it lost during this period. To do this we used our upper limit of $2 \times 10^{27}$ molecules/s for the last observation at 4 AU post-perihelion, and interpolated water production rates between $r_h$ = 2.3 and 4.0 AU to improve the computational accuracy of the numerical integration. Integrating the water production rate over the time it was measured by Swift (i.e. $t_p$ - 244 to +302 days, or $r_h$ = -3.4 to +4.0 AU) we find a total of $1 \times 10^{36}$ molecules, equivalent to $3 \times 10^{10}$ kg of water.

Integrating the dust production rate over the same period we find a total dust mass loss of $4 \times 10^{11}$ kg. Adding the dust and water, $4.3 \times 10^{11}$ kg of comet material were lost during the comet's pass through the inner solar system. Assuming a radius of ~ 5km and a density of 500 kg m$^{-3}$, the mass lost would correspond to the erosion of a layer of ~2 meters thick over the entire surface of the comet. This is significantly more than the thermal skin depth of the nucleus (centimeters) but less than the depths sampled by the Deep Impact experiment (few tens of meters; A'Hearn 2008).

### 5.4 Did CO$_2$ Drive the Icy Grains?

In Hartley 2, CO$_2$ drove icy grains into the coma (A'Hearn et al. 2011; Kelley et al. 2012). For Garradd, the only direct measurement of the CO$_2$ production rate was acquired by the Deep Impact spacecraft at 2 AU post-perihelion (Feaga et al. 2014). Line ratios of [OI] may be used to determine the column density ratios of H$_2$O, CO$_2$, and CO (Delsemme 1980; Festou & Feldman 1981  Bhardwaj & Raghuram 2012). However, the release rates of [OI] in the relevant states as well as physical processes in the coma that may affect these lines are not yet completely understood, resulting in large uncertainties in the derived abundances (McKay et al. (2012), DeCock et al. 2013, and references therein).

Assuming theoretical reaction rates by Bhardwaj & Raghuram (2012) we used the VLT/UVES measurements by DeCock et al (2013) to derive relative coma abundances of $CO_2/H_2O$ of 34% at -3.25 AU, 18% at -2.9 AU, 10% at -2.5 AU, and finally, around 5% at -2.08 AU. Based on [OI] observations at Apache Point Observatory, McKay et al. (2012) suggest constant $CO_2$ abundances of order 10-15% within 2 AU pre- and post-perihelion. Finally, the Deep Impact spacecraft measured a $CO_2/H_2O$ abundance of 8% at 2 AU post-perihelion (Feaga et al. 2014). It is of note that the [OI] observations were extracted from a narrow slit sized 0.44" x 12" (1,117 – 475 km at the comet) which implies that the [OI] measurements sample mostly gas directly sublimating from the nucleus.

Despite the large uncertainties in the abundances derived from [OI] observations, the overall trend of $Q(CO_2)/Q(H_2O)$ indicates a steep decrease before $t_p$ - 100 and constant levels or perhaps a slight recovery after perihelion. From Fig. 4, the contribution of the extended source to the water production rate peaked 50 days after the rapid decrease of the $CO_2/H_2O$ ratio. Given the large heliocentric distance of the comet (>2AU), this lag might be the result of the relatively long lifetimes (and low sublimation rates) of grains at those distances.

Should we expect CO to drive water ice after perihelion? Our results show no evidence of an excess source of water when the CO dominated the comet's gas production, suggesting that although the CO appears to drag dust along (Sec. 5.2), it is not a significant driver of ice.

## 6. SUMMARY AND CONCLUSIONS

Garradd is the first comet for which production rates of all three main volatiles ($H_2O$, CO, and $CO_2$) were measured, or at least inferred, during a significant part of its passage through the inner solar system. We demonstrate that the activity of Garradd was complex and changed significantly during the time it was observed. The orbital dynamics indicate that Garradd is not dynamically new but likely is young. The early outburst pre-perihelion, the presence of a cloud of icy grains, and the strange asymmetry of the CO production rate around perihelion are however more typically associated with dynamically new comets. This suggests that Garradd is 'dynamically young', i.e. in a transitional phase between 'new' and 'evolved' comets.

1. The total water production rate showed a very steep increase before perihelion ($\sim r^{-6}$), peaked 100 days before perihelion, and after remaining at a more or less constant level it decreased at a lower rate ($\sim r^{-4}$) after perihelion. Adding the dust and water production rates, $4.3 \times 10^{11}$ kg were lost during the comet's pass through the inner solar system, corresponding to a layer of one or two meters deep over the whole nucleus.

2. A constant active area of approximately 75 $km^2$ can explain the water sublimation from the nucleus. As Garradd had a nucleus of radius < 5.6 km (Boissier et al. 2013) it must have had a very high active fraction (>20%).

3. Between $t_p$ – 200 **d** ($r_h$ = 3 AU) and perihelion ($r_h$ = 1.55 AU), $H_2O$ was predominantly produced in the coma, likely by the sublimation of icy grains. The extended source may have been responsible for 75% of the water in the coma at its peak production.

4. At 4 AU pre-perihelion, Garradd had one of the highest dust-to-gas ($Af\rho/Q(OH)$) ratios ever observed in a comet, but it had typical dust-to-gas ratios around perihelion. Changes in the morphology of the dust distribution indicate the comet's dust production may be due to an outburst ending ~6 months before perihelion.

5. The high dust-to-gas ratio post-perihelion might be due to elevated CO production.

## **Acknowledgements**

We would like to thank Margaret Chester and Michael Siegel at the Swift Mission Operations Center for all of their help in scheduling and running the observation campaign of comet Garradd. DB and TLF received support through the Swift Guest Investigator program (Swift Cycles 7 and 8).

**Figures and Tables**

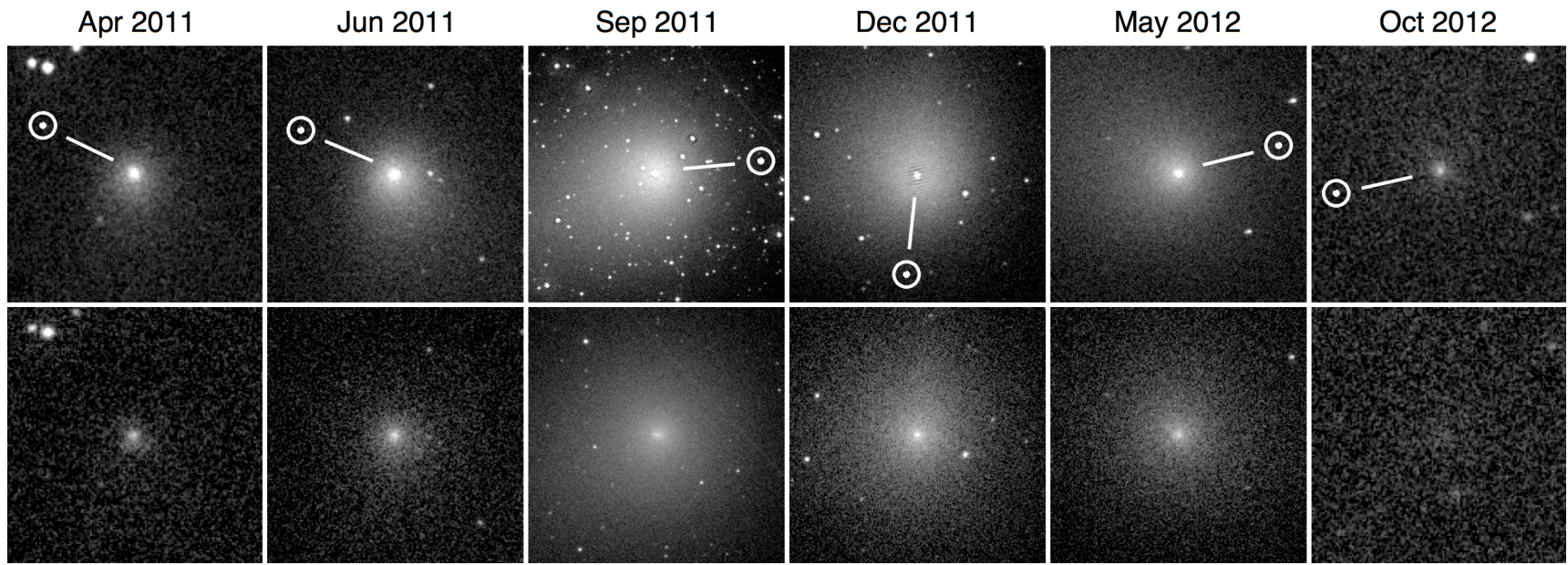

**Fig. 1** – A comparison of comet Garradd in raw images obtained with the V filter (top row) and the UVW1 filter (bottom) for the given dates of observation. The comet's coma tends to extend in the anti-solar direction in the V images, while it appears nearly spherical in the UVW1 images. Each panel has a field of view of $4.6 \times 10^5$ km, with North up, East to the left, and the Sun direction indicated.

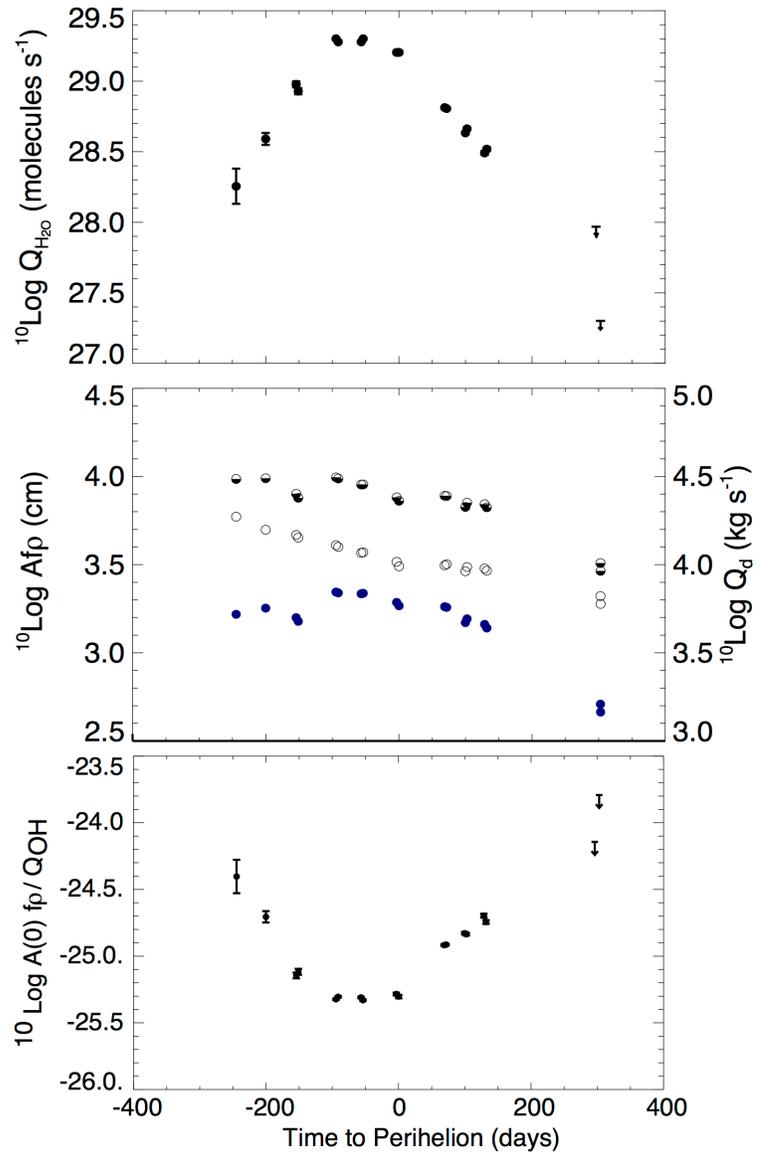

**Fig. 2 – Top:** $H_2O$ production rates from Swift/UVOT. **Middle:** Swift/UVOT V-band derived $Af\rho$ (open circles), phase function corrected $A(0)f\rho$ (half filled circles), and dust production rates (blue filled dots, secondary y-axis; shifted downward by 0.5 for clarity). Error bars in the $Af\rho$ data are smaller or comparable in size to the symbols used. **Bottom:** The ratio between phase corrected $A(0)f\rho$ and OH production rates Q(OH).

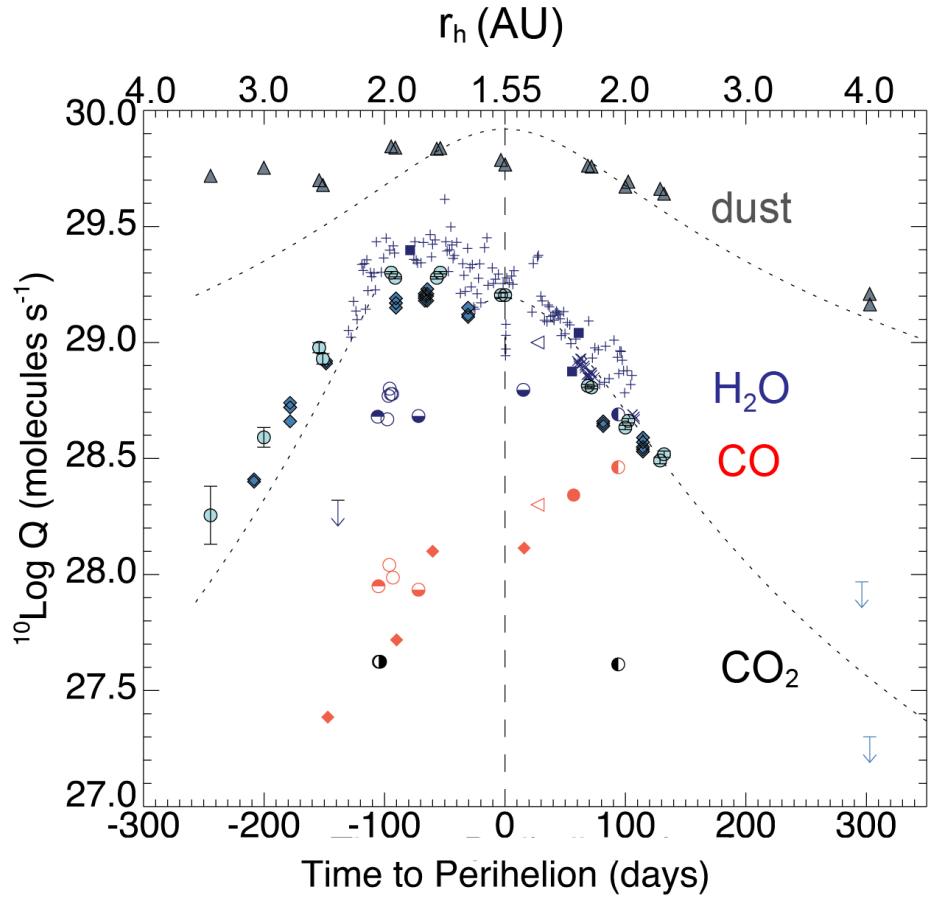

**Fig. 3.** – Temporal evolution of the production rates of the main cometary volatiles. Dotted lines through $H_2O$ values are drawn to guide the eye and indicate relations of $r^{-6}$ (total $H_2O$ pre-perihelion), $r^{-2}$ (dust production rate), and $r^{-4}$ (total $H_2O$ post-perihelion). **Dust production rates**: Swift/UVOT V-band (this paper; grey upward triangles; multiplied by $10^{26}$). **$H_2O$ production rates**: Swift/UVOT (light blue filled circles and arrows indicating upper limits); Lowell (Farnham et al. 2013; light blue diamonds); SOHO/SWAN (Combi et al. 2013; blue plusses); VLT/CRIRES (Paganini: et al. 2012; blue open circles and downward arrow pre-perihelion); Herschel/HIFI (Bockelée-Morvan et al. 2012, 2014; blue filled squares); Keck/NIRSPEC (Disanti et al. 2014; blue lower half filled circle); Keck/NIRSPEC & IRTF/CSHELL (Villanueva et al. 2012; blue upper half circle); DIF/MRI (Farnham in prep.; blue crosses); DIF/HRI-IR (Feaga et al. 2014; blue left half circle). HST (Feldman et al. ACM 2012; Blue left open triangle). **CO production rates**: JCMT (Yang ACM 2012; Feaga et al. 2014; red filled diamonds); VLT/CRIRES (Paganini et al. 2012, red open circles); Keck/NIRSPEC (Disanti et al. 2014; red lower half filled circle); red lower half filled circle); Keck/NIRSPEC & IRTF/CSHELL (Villanueva et al. 2012; red upper half circle); HST (Feldman et al. ACM 2012; left open red triangle); CSO and IRAM (Biver et al. ACM 2012; red filled circle); DIF/HRI-IR (Feaga et al. 2014; red left half circle). **$CO_2$ production rates**: relative $CO_2/H_2O$ abundances from VLT/UVES (Decock et al. 2013) combined with $H_2O$ production rates from VLT/CRIRES (Paganini et al. 2012); black right half circle; DIF/HRI-IR (Feaga et al. 2014; left filled half circle).

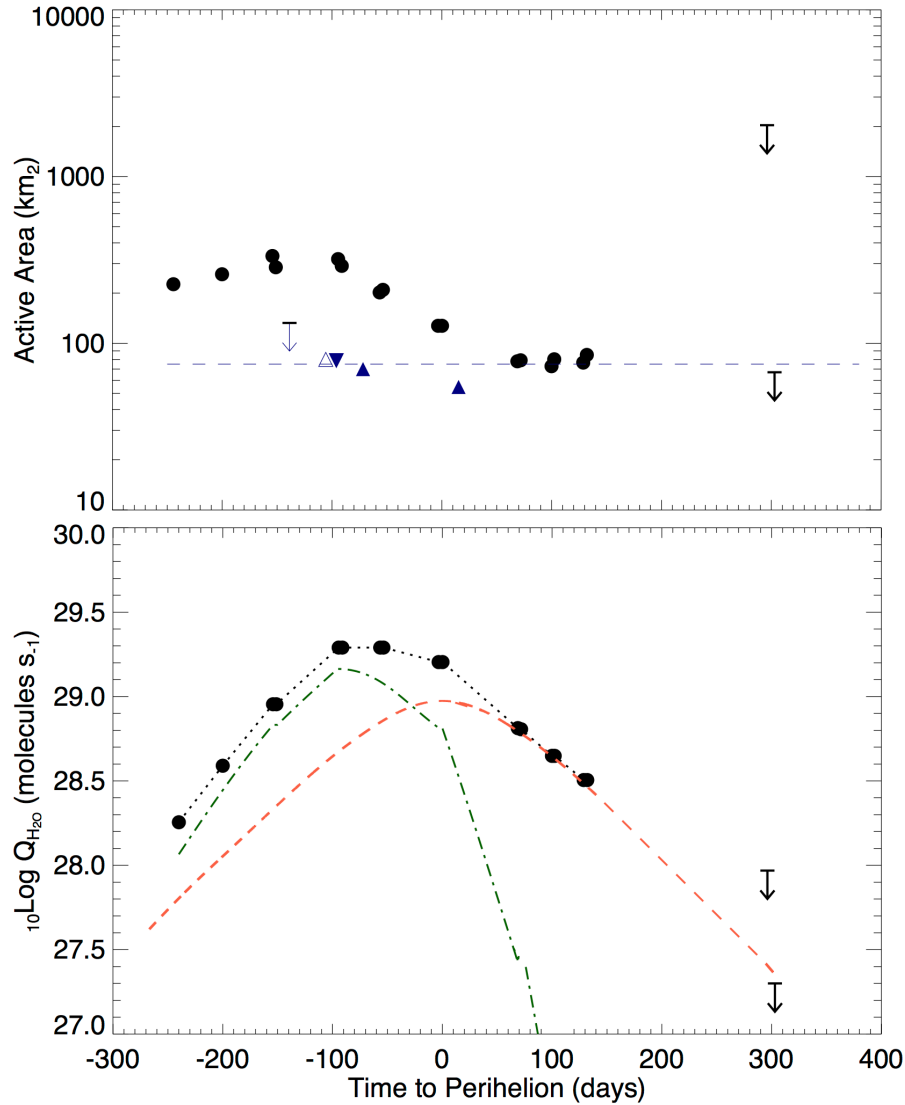

**Fig. 4.** – **Top**: Active areas for $H_2O$ (filled circles and upper limits at $t_p + 300d$), calculated using our production rates and assuming a local thermal equilibrium model. Measurements using small slits centered on the nucleus are shown in blue (upper limit pre perihelion and open upward triangle: Paganini et al. 2012; downward filled triangle: Villanueva et al. 2012; upward filled triangles: DiSanti et al. 2014), and may distinguish between $H_2O$ emanating from the nucleus and that produced by an extended source. The dashed blue line indicates a constant area of 75 $km^2$ for that scenario. **Bottom**: Contribution to the total water production (black dots; Swift/UVOT) by an active area on the nucleus of 75 $km^2$ (red dashed line) and sublimating grains in the coma (green dash-dotted line). The upper limit at $t_p = 296$ d was derived from the total flux measured within the UVW1 filter because no contemporaneous continuum image was acquired. On $t_p + 303$ d, both filters were used and we derived a much more constrained 3-sigma upper limit.

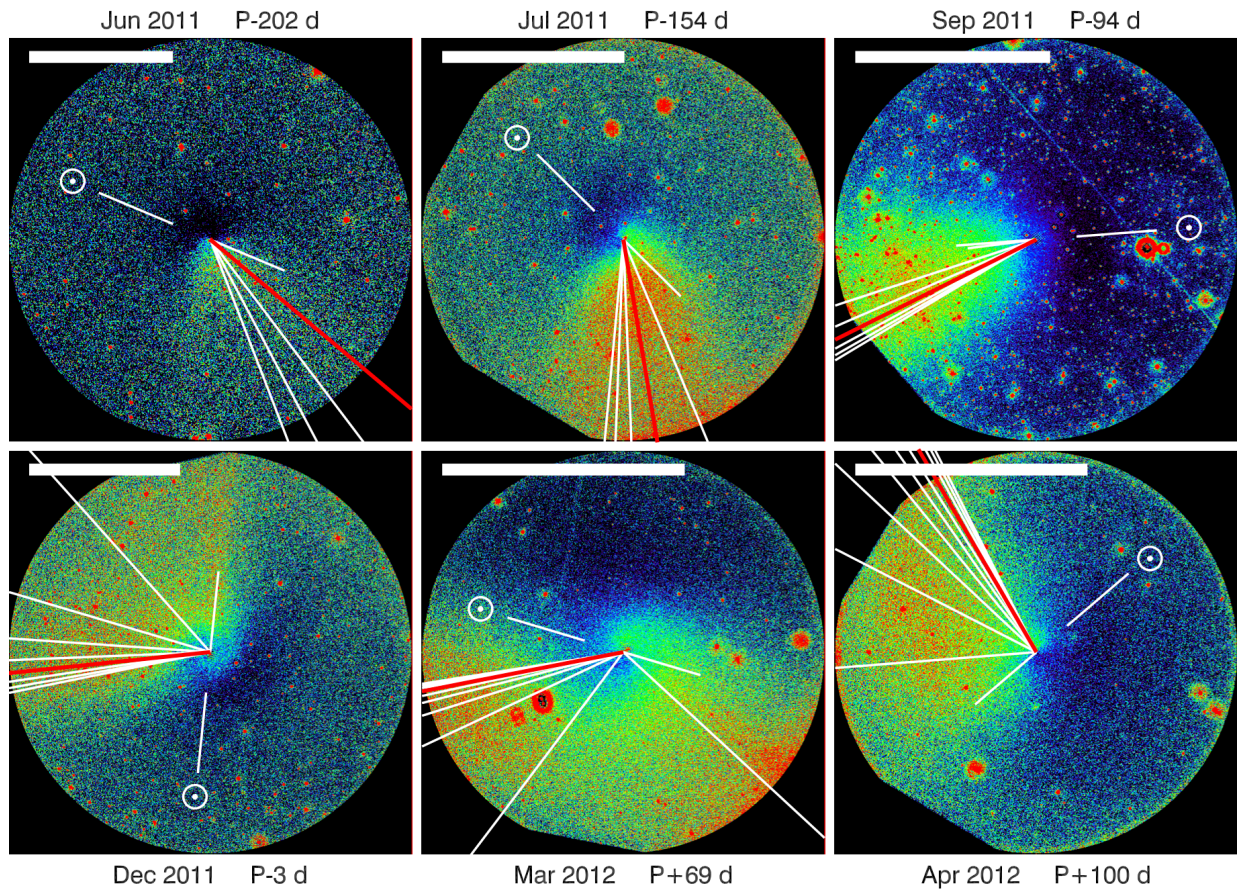

**Fig. 5** – Images of the dust tail of comet Garradd revealing times of increased dust emission. Images have been enhanced by dividing out the azimuthally averaged radial profile to improve the contrast of dust density in the tail. Synchrones are overplotted, starting 360 days before perihelion and stepping at intervals of 40 days up to the time of the observation, designated by the short synchrone in the anti-sunward direction. (For clarity, the 40-day synchrone closest to the observation time is sometimes left out.) The red highlighted synchrone denotes dust emitted at $t_p$ - 240 days. North is up, East is to the left, and the white bars represent a scale of 5 x $10^5$ km at the comet.

**Table 1** – Observing log of Swift/UVOT observations of Comet C/2009 P1 (Garradd). All errors and upper limits are 3-σ stochastic errors and do not include systematic uncertainties. V-band measurements listed in this table were measured from a fixed aperture of radius $5 \times 10^4$ km at the comet. All UVW1 fluxes and corresponding continuum V-band measurements were extracted from apertures of radius 50 arcsec, with the exception of the last three observations, for which we used an aperture of 25 arcsec in radius. †) $F_{OH}$ is the residual flux after continuum removal and corrected for filter transmission at the wavelength of the OH emission. ††) A(0)fρ/v is scaled to a phase angle of 0 degrees, and weighed by $r_h^{-0.5}$ to account for the heliocentric variation of the bulk outflow velocity. *) Upper limit based on flux in UVW1, no continuum subtracted. §) Flux and production rate derived from stacked images. In those cases, average values for midtime and observing geometry are given.

| Midtime /v(UT) | $\Delta T_{peri}$ (days) | Band | $T_{exp}$ (s) | $r_h$ (AU) | $dr_h$ (km/s) | $\Delta$ (AU) | Flux (W/m²) | $F_{OH}$ †) (W/m²) | $\Delta F$ (W/m²) | $m_v$ | $N_{mol}$ ($10^{32}$) | r FOV ($10^4$ km) | $Q_{H2O}$ ($10^{28}$/s) | $\Delta Q$ ($10^{28}$/s) | Afρ (cm) | $\Delta$Afρ (cm) | Phase (deg.) | Phase Corr 0 deg | A(0)fρ/ v ††) (cm) | $\Delta$A(0)f (cm) |
|---|---|---|---|---|---|---|---|---|---|---|---|---|---|---|---|---|---|---|---|---|
| 2011-04-23.261 | -244.4 | V | 260 | 3.42 | -16.9 | 3.95 | 1.8E-3 | | 3E-5 | 12.6 | | 4.9 | | | 5918 | 120 | 13.38 | 0.61 | 5229 | 105 |
| 2011-04-26.780 | -240.9 | UVW1 | 503 | 3.39 | -16.9 | 3.86 | 8.1E-4 | 8.1E-5 | 6E-5 | | 10 | 14 | 1.8 §) | 0.6 | | | 13.38 | | | |
| 2011-04-30.293 | -237.4 | V | 319 | 3.35 | -16.9 | 3.78 | 2.1E-3 | | 2E-5 | 12.5 | | 4.9 | | | 5791 | 96 | 14.71 | 0.59 | 5331 | 90 |
| 2011-06-04.680 | -202.0 | V | 319 | 3.01 | -16.9 | 2.86 | 3.8E-3 | | 3E-5 | 11.8 | | 5.0 | | | 4936 | 54 | 19.70 | 0.52 | 5519 | 63 |
| 2011-06-06.200 | -200.1 | UVW1 | 552 | 2.99 | -16.9 | 2.81 | 1.3E-3 | 5.2E-4 | 6E-5 | | 15 | 10 | 3.9 §) | 0.4 | | | 19.70 | | | |
| 2011-06-08.428 | -198.2 | V | 319 | 2.97 | -16.9 | 2.76 | 4.3E-3 | | 3E-5 | 11.7 | | 4.9 | | | 4989 | 51 | 19.96 | 0.51 | 5667 | 57 |
| 2011-07-22.436 | -154.2 | V | 230 | 2.55 | -16.5 | 1.69 | 1.4E-2 | | 9E-5 | 10.4 | | 5.0 | | | 4660 | 30 | 15.37 | 0.58 | 5003 | 33 |
| 2011-07-22.438 | -154.2 | UVW1 | 224 | 2.55 | -16.5 | 1.69 | 3.9E-3 | 2.5E-3 | 1E-4 | | 19 | 6.1 | 9.5 | 0.6 | | | 15.37 | | | |
| 2011-07-25.513 | -151.2 | V | 230 | 2.52 | -16.5 | 1.64 | 1.5E-2 | | 9E-5 | 10.4 | | 4.9 | | | 4497 | 30 | 14.57 | 0.59 | 4769 | 30 |
| 2011-07-25.515 | -151.2 | UVW1 | 226 | 2.52 | -16.5 | 1.64 | 4.0E-3 | 2.4E-3 | 1E-4 | | 16 | 5.9 | 8.5 | 0.6 | | | 14.57 | | | |
| 2011-09-20.243 | -94.4 | UVW1 | 414 | 2.01 | -14.2 | 1.57 | 1.2E-2 | 1.6E-2 | 2E-4 | | 40 | 5.7 | 20 | 0.3 | | | 29.56 | | | |
| 2011-09-21.114 | -93.6 | V | 370 | 2.00 | -14.1 | 1.58 | 2.3E-2 | | 9E-5 | 9.9 | | 4.9 | | | 4075 | 15 | 29.77 | 0.41 | 6995 | 27 |
| 2011-09-23.314 | -91.4 | V | 236 | 1.98 | -14.0 | 1.61 | 2.2E-2 | | 1E-4 | 9.9 | | 4.9 | | | 3987 | 21 | 30.26 | 0.41 | 6915 | 36 |
| 2011-09-23.583 | -91.1 | UVW1 | 445 | 1.98 | -13.9 | 1.61 | 1.2E-2 | 1.6E-2 | 2E-4 | | 41 | 5.8 | 19 | 0.3 | | | 30.31 | | | |
| 2011-10-28.080 | -56.6 | V | 230 | 1.73 | -10.4 | 1.98 | 1.8E-2 | | 1E-4 | 10.2 | | 4.9 | | | 3687 | 21 | 30.09 | 0.41 | 6836 | 39 |
| 2011-10-28.083 | -56.6 | UVW1 | 186 | 1.73 | -10.4 | 1.98 | 1.5E-2 | 2.2E-2 | 3E-4 | | 58 | 7.2 | 19 | 0.3 | | | 30.09 | 0.41 | | |

| Date | | | | | | | | | | | | | | | | | | | | |
|---|---|---|---|---|---|---|---|---|---|---|---|---|---|---|---|---|---|---|---|---|
| 2011-12-23.855 | 0.18 | v | 96 | 1.55 | 0.04 | 2.01 | 1.8E-2 | | 2E-4 | 10.1 | | 5.0 | | | 3093 | 27 | 28.54 | 0.42 | 5846 | 54 |
| 2011-12-23.856 | 0.18 | uvw1 | 61 | 1.55 | 0.04 | 2.01 | 1.2E-2 | 1.4E-2 | 4E-4 | | 50 | 7.3 | 16 | 0.6 | | | 28.54 | | | |
| 2012-03-01.448 | 68.8 | v | 260 | 1.81 | 11.9 | 1.27 | 3.4E-2 | | 1E-4 | 9.5 | | 5.0 | | | 3135 | 12 | 31.75 | 0.40 | 5775 | 24 |
| 2012-03-01.451 | 68.8 | uvw1 | 242 | 1.81 | 11.9 | 1.27 | 1.1E-2 | 1.4E-2 | 2E-4 | | 10 | 4.6 | 6.5 | 0.09 | | | 31.75 | | | |
| 2012-03-04.453 | 71.8 | v | 289 | 1.83 | 12.2 | 1.27 | 3.4E-2 | | 1E-4 | 9.5 | | 5.0 | | | 3179 | 12 | 30.99 | 0.41 | 5728 | 21 |
| 2012-03-04.456 | 71.8 | uvw1 | 258 | 1.83 | 12.2 | 1.27 | 1.8E-2 | 1.4E-2 | 2E-4 | | 9.9 | 4.6 | 6.4 | 0.09 | | | 30.99 | | | |
| 2012-04-01.569 | 99.9 | v | 230 | 2.05 | 14.5 | 1.53 | 1.7E-2 | | 1E-4 | 10.2 | | 5.0 | | | 2901 | 18 | 27.81 | 0.43 | 4690 | 30 |
| 2012-04-01.571 | 99.9 | uvw1 | 191 | 2.05 | 14.5 | 1.53 | 6.8E-3 | 8.0E-3 | 2E-4 | | 8.2 | 5.5 | 4.3 | 0.09 | | | 27.81 | | | |
| 2012-04-04.044 | 102.4 | v | 230 | 2.07 | 14.7 | 1.57 | 1.6E-2 | | 9E-5 | 10.3 | | 5.0 | | | 3067 | 18 | 27.76 | 0.43 | 4928 | 30 |
| 2012-04-04.046 | 102.4 | uvw1 | 185 | 2.07 | 14.7 | 1.57 | 6.8E-3 | 8.2E-3 | 2E-4 | | 9.1 | 5.7 | 4.6 | 0.09 | | | 27.76 | | | |
| 2012-04-30.445 | 128.8 | v | 260 | 2.31 | 15.9 | 2.18 | 6.8E-3 | | 6E-5 | 11.2 | | 5.0 | | | 3010 | 30 | 25.77 | 0.43 | 4585 | 45 |
| 2012-04-30.448 | 128.8 | uvw1 | 260 | 2.31 | 15.9 | 2.18 | 3.4E-3 | 3.5E-3 | 1E-4 | | 9.6 | 7.9 | 3.1 | 0.09 | | | 25.77 | | | |
| 2012-05-03.730 | 132.1 | v | 244 | 2.34 | 16.0 | 2.26 | 6.0E-3 | | 6E-5 | 11.4 | | 4.9 | | | 2922 | 30 | 25.28 | 0.44 | 4372 | 45 |
| 2012-05-03.733 | 132.1 | uvw1 | 240 | 2.34 | 16.0 | 2.26 | 3.2E-3 | 3.5E-3 | 1E-4 | | 11 | 8.2 | 3.3 | 0.12 | | | 25.27 | | | |
| 2012-10-14.835 | 296.2 | uvw1 | 452 | 3.92 | 16.6 | 4.50 | 8.3E-5 | <1.7E-4 | | | <6.1 | 8.2 | <0.9 *) | | | | 11.08 | 0.66 | | |
| 2012-10-21.248 | 302.6 | v | 372 | 3.98 | 16.5 | 4.48 | 3.7E-4 | | 2E-5 | 14.4 | | 4.9 | | | 2099 | 96 | 11.75 | 0.65 | 1618 | 75 |
| 2012-10-21.516 | 302.8 | uvw1 | 1156 | 3.98 | 16.5 | 4.48 | 6.7E-5 | 0 | 7.7E-6 | | <0.3 | 8.1 | <0.2 § | | | | 11.75 | | | |
| 2012-10-21.515 | 302.8 | v | 379 | 3.99 | 16.5 | 4.48 | 3.3E-4 | | 6E-6 | 14.5 | | 4.9 | | | 1893 | 96 | 11.75 | 0.65 | 1459 | 75 |